\pdfoutput=1

\documentclass[aip,jmp,amsmath,amssymb,reprint]{revtex4-1}

\usepackage{graphicx}
\usepackage{dcolumn}
\usepackage{bm}
\usepackage[caption=false]{subfig}
\usepackage{textcomp}
\usepackage{booktabs}
\usepackage{hhline}
\begin{document}

\title[]{Modeling Electronic Excited States of Molecules in Solution}

\author{Tim J. Zuehlsdorff}
\email{tzuehlsdorff@ucmerced.edu}
\affiliation{Chemistry and Chemical Biology, University of California Merced, N. Lake Road, CA 95343, USA}
\author{Christine M. Isborn}
\email{cisborn@ucmerced.edu}
\affiliation{Chemistry and Chemical Biology, University of California Merced, N. Lake Road, CA 95343, USA}
\date{\today}

\begin{abstract}
The presence of solvent tunes many properties of a molecule, such as its ground and excited state geometry, dipole moment, excitation energy, and absorption spectrum. Because the energy of the system will vary depending on the solvent configuration, explicit solute-solvent interactions are key to understanding solution-phase reactivity and spectroscopy, simulating accurate inhomogeneous broadening, and predicting absorption spectra. In this tutorial review, we give an overview of factors to consider when modeling excited states of molecules interacting with explicit solvent. We provide practical guidelines for sampling solute-solvent configurations, choosing a solvent model, performing the excited state electronic structure calculations, and computing spectral lineshapes. We also present our recent results combining the vertical excitation energies computed from an ensemble of solute-solvent configurations with the vibronic spectra obtained from a small number of frozen solvent configurations, resulting in improved simulation of absorption spectra for molecules in solution. 
\end{abstract}

\maketitle

The presence of solvent around a molecule affects its energy, properties, dynamics, and reactivity, in both the ground and excited state. Because many excited state phenomena of chemical interest happen in solution and in complex interfacial environments, including photosynthesis, photocatalysis, photo-induced charge and proton transfer, and fluorescence in biological systems, it is important to be able to model these excited state reactions while accurately simulating the solvent environment. In addition, both static and time-resolved absorption and fluorescence spectroscopies serve as useful tools to elucidate chemical reactions and pathways, and simulation of these solution-phase spectra can help to achieve understanding of the electronic rearrangements governing chemical reactions. 

If theoretical models are to provide reliable simulations of solution-phase chemistry, accurate models of the solvent environment are required. 
The solvent environment can have a strong influence on an absorption spectrum due to polarization and direct solute-solvent interactions such as hydrogen bonding. Therefore, the modeling of absorption spectra of molecules in solution is often considered a vital step in developing and benchmarking  methods to account for the complex solvent environment. 

\begin{figure}
    \centering
    \includegraphics[width=0.48\textwidth]{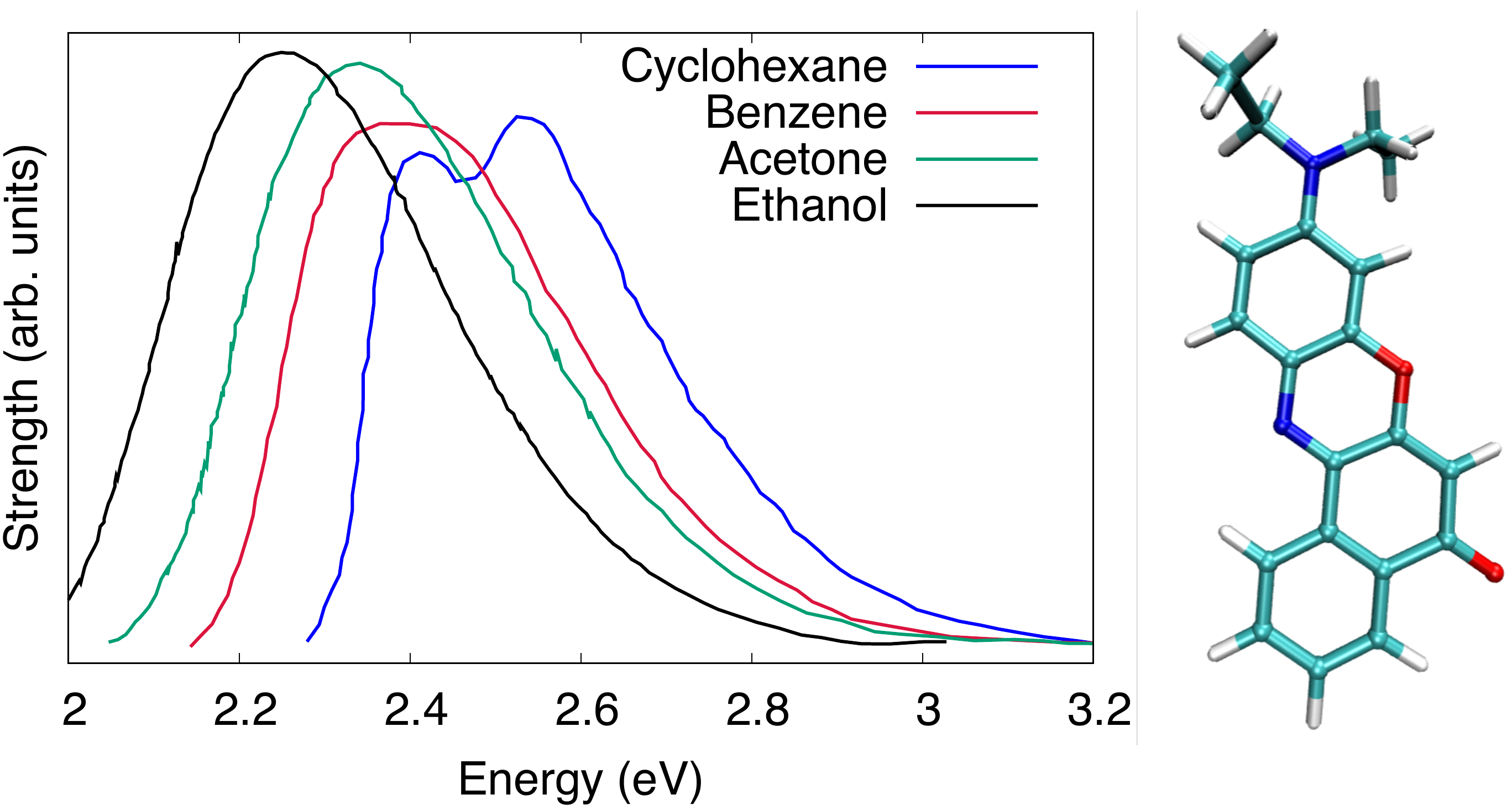}
    \caption{Experimental absorption spectra for Nile red in four solvents. The observed solvatochromic red-shift approximately correlates with increasing solvent polarity ($\epsilon_0$=2.0, 2.3, 20.7, 24.5 for cyclohexane, benzene, acetone and ethanol respectively) and the spectral shapes also undergo significant changes. Although cyclohexane and benzene have very similar dielectric constants, they produce rather different spectral shapes, suggesting that explicit solute-solvent interactions play an important role in determining the lineshape. The experimental data in this plot was taken from Davis \emph{et al.}\cite{Davis_1966}}
    \label{fig:nile_red_experiment}
\end{figure}

Although continuum solvent approaches are computationally attractive because they use the bulk properties of the solvent to polarize a solute, and thus usually do not sample solute-solvent configurations, in reality it is specific solvent configurations rather than a solvent continuum that determine excited state properties. Also, the simulation of spectral lineshapes, such as those shown in Fig.~\ref{fig:nile_red_experiment} that include both the high-energy tail due to vibronic transitions and the inhomogeneous broadening due to solute-solvent interactions, poses a significant challenge for continuum methods. Thus, here we concentrate our discussion on explicit solvent, with a particular focus on treating the solvent quantum mechanically. Previous excellent reviews and perspectives have focused on continuum\cite{Tomasi_2005,Cramer_1999, Mennucci_2012, Mennucci_2015, Lipparini_2016} or fragment \cite{DeFusco_2011, Gordon_2012} approaches for modeling solvent and we include a brief overview of these methods, along with some other solvent models, in this tutorial review. 

Modeling excited states with explicit quantum mechanical (QM) solvent has a number of advantages over continuum approaches. When treating the solvent with QM, the solute and solvent can undergo mutual polarization as well as transfer of charge. The amount of charge transfer from solute to solvent may be quite large in some cases (see, for example, Ref. \cite{Ufimtsev_2011}) and the degree of charge transfer from the solute to the solvent may differ when going from the ground to the excited state.\cite{Provorse2016} The large QM regions of solvent, approximately 1-2 solvation shells, that are required for convergence of excitation energies \cite{Isborn2012, Zuehlsdorff2016, Provorse2016, Milanese2017}, indicate that specific interactions with explicit solvent are quite long-ranged. When performing ab initio molecular dynamics (AIMD) or ab initio geometry optimizations with explicit solvent, the system may also undergo proton or atom transfer between the solute and solvent. Treating the solvent with QM is essential for modeling processes such as excited state proton transfer from solute to solvent, as was recently done for the Lewis photoacid methyl viologen.\cite{Hohenstein_2016} 

\begin{figure*}
    \centering
    \includegraphics[width=0.95\textwidth]{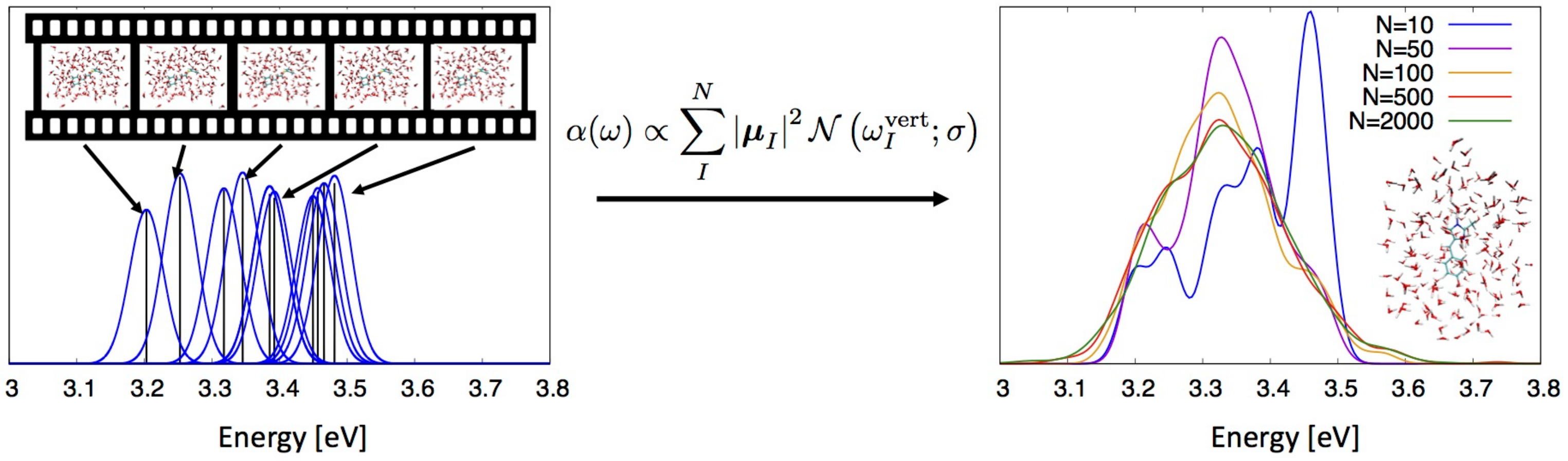}
    \caption{To sample an ensemble of solute-solvent configurations for simulation of an absorption spectrum, structures obtained from snapshots of a molecular dynamics trajectory are used for the computation of vertical excitation energies. Spectra are constructed by averaging over the vertical excitations calculated for different solute-solvent snapshots. In the ensemble approach, the $\delta$-functions corresponding to individual vertical excitations are convoluted with a Gaussian function $\mathcal{N}$ of width $\sigma$ centered at $\omega^{vert}$. The figure on the left shows the vertical excitation energies $\omega^{vert}$ and corresponding oscillator strengths as a stick spectrum for N=10 snapshots, with each stick dressed with a Gaussian function. The sum of the Gaussian functions creates an absorption spectrum within the ensemble approach. The figure on the right shows the convergence of the ensemble spectrum computed for the GFP chromophore anion in water with respect to the number of snapshots sampled. }
    \label{fig:convergence_num_snapshots}
\end{figure*}

In concert with potential increased accuracy, performing excited state calculations with QM explicit solvent presents a number of challenges. Some of these challenges originate from the choice of electronic structure method or the choice of classical method for surrounding the QM region; we discuss these issues later in this tutorial review. The primary challenge to overcome is the increase in computational cost from the larger QM region and from the need to sample solute-solvent configurations. Over the past decade, exited state linear-scaling electronic structure algorithms \cite{Zuehlsdorff_2013, Zuehlsdorff_2015} and the development of electronic structure code to take advantage of stream-processing computer hardware \cite{Isborn_2011} has led to the advent of large-scale excited state calculations. These large-scale QM calculations hopefully will not only lead to more accurate simulations by allowing the QM treatment of solvent, but can also provide insight into the nature of solvation and the role that solvent plays in tuning excitation energies.

In this tutorial review we give an overview of the current state of the computation of vertical excitation energies and absorption spectra in solution, with a focus on including explicit solvent in the calculations. We first examine how to sample configurations of a molecule interacting with explicit solvent. Next we give a brief introduction to different solvation models, both continuum and explicit, followed by issues to consider regarding various excited state electronic structure methods. We lastly discuss simulating absorption lineshapes for molecules in solution, examining solvation effects and vibronic effects. 

\section{Sampling of solute-solvent configurations}

Most excited state calculations are carried out on an optimized structure. If trying to include specific solvent effects on the structure and excited state, it might be possible to find an optimized geometry for the solute and solvent if a small number of solvent molecules are included in the calculation. However, optimizing the geometry of a solute-solvent system is often challenging because of low-frequency degrees of freedom. Furthermore, once the number of explicit solvent molecules becomes large, a relaxed structure obtained through a geometry optimization is likely no longer unique, as the full potential energy surface of the system will contain a large number of minima with similar energy, corresponding to different solvent arrangements. Another complication, if comparing to data from an experiment, is that a fully optimized structure will not include temperature effects and the room temperature solvent structure might vary considerably from the optimized structure. 
 
To include temperature effects as well as to sample a broad distribution of specific solute-solvent configurations while accounting for the vibrational motion of the system on an anharmonic potential energy surface, a molecular dynamics (MD) simulation can be performed to obtain many ‘snapshots’ from the MD trajectory at the desired temperature. For a solution-phase spectrum that can be compared to experiment, excited state calculations should be performed for many solute-solvent configurations. For modeling absorption, the MD simulates the solute in the ground state. For modeling fluorescence, the MD simulates the solute in the excited state with the solvent in equilibrium with with this excited state structure. It is often challenging to obtain computationally affordable excited state gradients, therefore fluorescence simulations with explicit solvent are currently relatively rare. 

Most MD simulations treat the nuclei classically, resulting in a classical Boltzmann ensemble of configurations. If a quantum ensemble of configurations is desired, nuclear quantum effects can be included in the ensemble by using a harmonic-oscillator Wigner distribution\cite{Barbatti2016, Nogueira2018} or by performing path integral MD (PIMD).\cite{Markland2018} Because these nuclear quantum effects include vibrational zero-point energy, the sampled configurations are higher on the potential energy surface, and, compared to purely classical configurations, simulated absorption spectra are often red-shifted. The higher energy PIMD configurations also tend to sample more anharmonic regions of the potential energy surface, resulting in broader absorption spectra. Recent studies of absorption spectra computed from PIMD configurations suggest that the majority of the spectral broadening is not due to nuclear quantum effects in the solvent, but is instead due to the increased bond lengths sampled by the solute.\cite{Law2015, Law2018}

Absorption spectra in solution are broader than absorption spectra in vacuum due to the inhomogeneous solvent environment. When modeling absorption spectra, sampling over many solute-solvent configurations allows the simulation of this solvent-induced inhomogeneous broadening. With explicit solvent, there is no restriction on the shape of the broadening, i.e. it need not lead to a Gaussian distribution of excitation energies as is often assumed when modeling spectra with a continuum approach. The assumption of a Gaussian distribution of excitation energies from the addition of a solvent environment remains valid if the solute and solvent do not have specific interactions and the solvent provides only a weak electrostatic bath for the solute. If specific solute-solvent interactions are present, such as $\pi$-stacking or hydrogen-bonding, the distribution of excitation energies will no longer be Gaussian. Explicit solvent in the computation of vertical excitation energies can therefore yield more accurate spectral lineshapes for systems with strong solute-solvent interactions.  

A simulated spectrum that is converged with respect to the number of  solute-solvent configurations requires the calculation of vertical excitation energies for many snapshots. Although the specific number of snapshots required for convergence will depend on the system as well as the choice of a spectral broadening parameter, we have found that for solution-phase UV-Vis spectra, approximately 500 snapshots is appropriate. As an illustration, the absorption spectra computed by broadening the vertical excitation energies with a Gaussian function of width $\sigma$=0.021 eV for an ensemble of snapshots of the chromophore of green fluorescent protein (GFP) in water is shown in Fig.~\ref{fig:convergence_num_snapshots}. Spectra computed with N=10 or N=50 snapshots show significant deviation in shape and energy for the absorption maximum, whereas the spectrum computed with N=500 snapshots shows good agreement with the N=2000 snapshot spectrum. 

After solute-solvent configurations are obtained, the next step when simulating spectra in solution is to perform excited state calculations for each configuration. One is then faced with the choice of both the solvent model and the electronic structure method for describing the excited state. We discuss each of these below before going into more detail regarding the computation of spectral lineshapes. 

\section{Solvent models for excited state calculations}

Almost all solvent models account for polarization of the solute, and so can therefore capture some aspects of solvatochromism \cite{Buncel_1990, Reichardt_1994,Marini_2010}, as seen in Fig.~\ref{fig:nile_red_experiment}. If the excited state dipole moment of a solute is larger than the ground state dipole moment, the excited state will often be preferentially stabilized by the solvent polarization environment, leading to a spectral shift to lower energies (a red-shift, or bathochromic shift) with increasing solvent polarity. This is known as positive solvatochromism. If the ground state is preferentially stabilized over the excited state, a spectral shift to higher energies occurs (a blue-shift, or hypsochromic shift), known as negative solvatochromism. However, beyond polarization effects, specific solute-solvent interactions such as hydrogen-bonding can also affect the solvatochromic shift, and these interactions certainly play a large role in determining spectral shape. Therefore, for capturing the full change in energy, intensity, and line shape due to changes in the solvent environment, explicit solvent should be included in the theoretical model.

In polar solvents, the solute is polarized by the solvent environment, and the degree of energetic stabilization is often determined by the dipole moment of the solute and the dielectric constant of the solvent. For polar solvent molecules, the dipole moments of the solvent will align with the dipole moment of the solute to provide the best electrostatic configuration. Upon electronic excitation, there is often a change in the dipole moment of the solute, leaving the solvent no longer in an equilibrium configuration around the solute.
The electronic distribution of the solvent can change on the same attosecond timescale as the electronic rearrangement of the solute, but the nuclear rearrangement of the solvent molecules to stabilize the excited state occurs much more slowly, often on the picosecond to nanosecond timescale.

The solvent models used in the computation of excited states are the same solvent models used in the computation of ground states. We here give an overview of various solvent models that have been implemented within excited state electronic structure calculations. Sections 2.1 and 2.2 contain summaries of classical continuum solvation models and explicit solvation approaches, respectively, that are mainly intended for readers unfamiliar with the details of the different ways solvent effects are accounted for in practical calculations of electronic excitations in solution. Readers already familiar with these approaches might want to skip this part of the tutorial review and move to section 2.3, which contains a more detailed discussion of factors to consider when combining a quantum treatment of parts of the solvent environment with classical solvent models. 

\subsection{Continuum solvation models}

In continuum solvation models, the separation of electronic and nuclear rearrangement timescales is accounted for through the use of the static and optical (also called dynamical) dielectric constants of the solvent, denoted $\epsilon_0$ and  $\epsilon_\infty$, respectively. Ground state single point energy calculations, as well as both ground state and excited state geometry optimizations, are performed using $\epsilon_0$, whereas vertical excitation energies and non-equilibrium properties of the excited state are computed using $\epsilon_\infty$. The value of the static dielectric constant varies substantially going from nonpolar solvents such as hexane ($\epsilon_0$=1.88) to polar solvents such as water ($\epsilon_0$=78.36), whereas the optical dielectric constant is a measure of the polarizability rather than polarity of the solvent and can be quite similar for nonpolar and polar solvents ($\epsilon_\infty$=1.89 for hexane, $\epsilon_\infty$=1.78 for water). 

One of the simplest continuum, implicit solvent approaches is the Onsager model \cite{Onsager_1936}, in which the solvent response is computed by treating the solute as a simple dipole moment in a spherical cavity. More advanced continuum methods that use an apparent surface charge formulation are now implemented in most electronic structure packages. These methods place charges induced by the electrostatic potential of the solute on the surface of a molecular cavity often formed by atom-centered spheres with atom-specific radii or from the isodensity of the solute. \cite{Tomasi_2005,Cramer_1999, Mennucci_2012, Mennucci_2015, Lipparini_2016, Cances_1997, Barone_1998, Chipman_1999, Klamt_2011} Because the shape of the cavity and the proximity of the cavity surface to the QM electron density will alter the polarization environment, the choice of the cavity will affect the ground and excited state energies. A smaller cavity will allow the surface charges to be closer to the QM electron density, increasing the polarization and stabilization provided by the solvent environment. The continuum apparent surface charge equations are solved self-consistently to determine the mutual polarization of the solute and the surface charges representing the solvent. Some continuum methods include non-electrostatic contributions to the energy, such as dispersion and creation of the cavity, but these energy corrections do not change the electronic wavefunction, and therefore will not change the transition energies or transition dipole moments.

The vertical excitation energy can be determined with a continuum approach by allowing the solvent to polarize in response to the transition density or in response to the difference in the excited state and ground state electron densities. When using the transition density, the continuum equations are solved in conjunction with response theory and this is known as the linear response (LR) formulation. When using the difference density, the excited state electron density must be obtained, and the technique is referred to as the state-specific (SS) formulation. Studies have shown that the LR approach provides incorrect solvent polarization for transitions with a large change in electron density, such as charge-transfer transitions.\cite{Cammi_2005, Corni_2005, Caricato_2006,Improta_2006, Guido_2015, Mewes2017} 

Self-consistently solving for the electron density of the QM region and the response of the solvent increases the cost of the calculation compared to a vacuum calculation. However, the use of a continuum solvent model is more computationally affordable than using explicit QM solvent because, in practice, the continuum calculation is usually only performed for optimized geometries of the QM region and additional basis functions are not required for modeling the QM solvent molecules. Because the solvent dielectric constant is a bulk property, continuum approaches account for the average response of the solvent and there is no need to average over solute-solvent configurations. The other solvation models discussed here use an explicit representation of the solvent molecules, and therefore will require sampling of solute-solvent configurations in order to model bulk properties. 

\subsection{Explicit solvent models}

The disadvantage of continuum implicit solvent approaches is the lack of specific solute-solvent interactions. An alternative classical solvation model for electronic excited states that accounts for polarization and specific solute-solvent interactions is to electrostatically embed the molecular mechanical (MM) point charges of the solvent atoms into the one electron term in the QM Hamiltonian via the QM/MM approach.\cite{Lin2006, Senn2009} This charge distribution accounts for the electrostatic contributions of specific solute-solvent interactions, such as the electrostatic effects of a hydrogen-bond. Most MM force fields employ fixed point charges and are therefore non-polarizable; they polarize the QM region, but because the values of the point charges are fixed, there is no self-consistent (mutual) polarization between the QM region and the solvent. The solvent cannot polarize in response to the change in solute electron density upon excitation, and both the nuclear and electronic response of the solvent is then ‘frozen’ during the electronic excitation. Consequently, the MM point charges do not enter into the calculation of the excited state, except indirectly through polarization of the ground state.  

Solvent MM force fields are often parametrized to give agreement with bulk solvent properties. The resulting point charges may not be the ideal values to represent solvent interacting with a QM solute. Bare MM point charges may over-polarize the QM wavefunction, altering the excitation energy. Although techniques have been proposed that smear\cite{Das2002} and scale\cite{Vreven2006} the MM charges interacting with the QM region, these alterations can lead to incorrect long-range polarization. 

Including MM point charges around a QM region is usually more computationally affordable than using a continuum model because there is no self-consistent polarization between the QM region and the charges. Very large MM regions can easily be used because the computation of the one-electron integrals is not the bottleneck in QM calculations and therefore the number of MM point charges generally does not significantly affect the time for the QM calculation. 

Polarizable MM solvent includes both self-consistent polarization and specific solute-solvent interactions. These polarizable models often use induced dipoles, fluctuating charges, or a Drude oscillator to allow for a changing electronic environment.\cite{Bryce_1997, Lamoureux_2003, Jensen_2003, Lipparini_2011, Thiel_2012} When combined with excited state formulations, these methods account for the electronic rearrangement of the solvent in response to the excitation of the solute.\cite{Sneskov_2011, Steindal_2011, Zeng_2015, List_2016, Loco_2016}  

Fragment based methods\cite{Raghavachari2015, Collins2015} bridge the realms of MM and fully QM solvent models, both in terms of accuracy and computational cost. One example of a fragment based approach that has been interfaced with a variety of excited state electronic structure techniques is the effective fragment potential method.\cite{DeFusco_2011, Gordon_2012} Within this method, each solvent molecule interacts with the solute with electrostatic, polarization, dispersion, exchange-repulsion, and charge-transfer terms contributing to the interaction energy. The electrostatic potential contains terms from charges, dipoles, quadrupoles, and octopoles. Many-body effects are accounted for by self-consistently determining the polarization interaction between each solvent fragment and the QM solute as well as the other solvent fragments. The polarizable potential is created from polarizabilities and distributed multipoles obtained from ab initio computations. 

A related way to also account for specific solute-solvent interactions is through the use of an embedding potential.\cite{Wesolowski2015,Sun2016, Goez2018} Embedding methods allow a system treated with a high-level of theory to experience the potential of an environment treated with a lower level of theory, enabling the QM treatment of large amounts of solvent around a solute.\cite{Wesolowski1993}  The embedding approach has been extended to response theory,\cite{Casida2004, Gomes2012} allowing the calculation of excitation energies within the field of many solvent molecules. \cite{Neugebauer2005, Bennie2017}

\subsection{Combining QM explicit solvent with classical solvent models}

In addition to the solvent models described above, the solvent can also be treated on equal footing with the solute. Using the same level of QM electronic structure theory for the solute and solvent will increase the computational cost compared to the more approximate solvent models discussed previously. The full interaction between the solute and solvent can also complicate the excited state analysis because states may have contributions from both solute and solvent. However, the trade-off of these extra complications may be balanced with improved accuracy.

If the solvent is not treated separately from the solute, the solute and solvent undergo self-consistent polarization with inclusion of exchange and perhaps correlation and dispersion energies, as dictated by the electronic structure theory. Additionally, charge can transfer between the QM solvent and the QM solute. For non-polar solvents, dispersion interactions between the solute and solvent may be the substantial contributor to changes in the transition energies. These dispersion interactions can be easily included if the solvent is treated as QM and if the electronic structure method employed accounts for dispersion. The solute-solvent interactions may be significant at short-range, i.e. within the first solvation shell, but they fall off quickly with distance and a more approximate model for the solvent can be used for long-range interactions. 

\begin{figure}
\centering
{\includegraphics[width=0.48\textwidth]{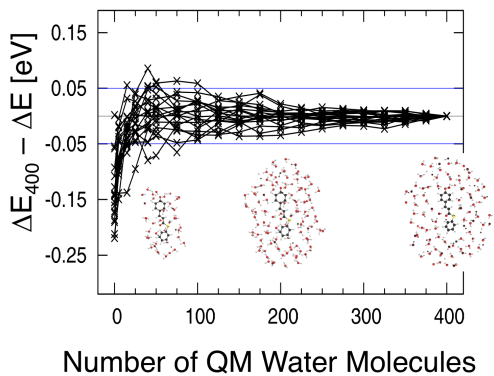}}
\caption{The deviation of the computed excitation energy compared to the value with 400 QM water molecules shows convergence within 0.05 eV when approximately one solvation shell is treated with QM (100 QM water molecules). The QM solvent is surrounded by MM solvent. The solute is the anionic chromophore of photoactive yellow protein. Figure is adapted with permission from Ref. \cite{Milanese2017}. Copyright 2017 American Chemical Society.}
\label{fig:ex_energy_convergence}
\end{figure}

Given that a QM treatment of solvent might be desirable, the careful researcher may next wonder how much solvent should be treated quantum mechanically. A number of studies have explored this question for different molecular properties,\cite{deLima2010, Ghosh2011, Flaig2012, Liao2013, Tentscher2015, Kulik2016} and the study of convergence of excitation energies with increasing size of molecular environment has become an area of active research.\cite{Murugan2010, Isborn2012, Ma2012, Eilmes2014, Provorse2016, Zuehlsdorff2016} Although in some systems a large amount of QM solvent may not be critical for accurately computing the property of interest, we have found that for excited states, values within approximately 0.05 eV of the converged value can be obtained with a full solvation shell treated with QM, as shown in Figure \ref{fig:ex_energy_convergence}.\cite{Milanese2017} The amount of solvent required for this level of accuracy was similar for both polar and less-polar solutes, although this was somewhat dependent on the choice of the electronic structure method. We also found that the time for computation could be lessened, while maintaining the accuracy of the computed excitation energy, by decreasing the size of the basis set for QM solvent molecules more distant from the solute. 

If the goal is to model the excited states of a molecule in solution rather than in a solvent cluster, the QM solvent should be surrounded by a more approximate solvent model. This environment will not only account for long-range solute-solvent electrostatic interactions, but will also improve the description of the QM solvent. For example, both MM point charges and a polarizable continuum model surrounding a QM solvent layer increased the band gap of the solvent compared to the vacuum solvent cluster, bringing it closer to the experimental value of bulk solvent.\cite{Isborn2013}  Rather counter-intuitively, we found that the addition of the classical environment decreased the computational cost of the calculations; this was due to the increase in the band gap leading to need for fewer self-consistent field iterations being necessary to converge the ground state electron density. When taking a subset of solvent around a molecule from an MD simulation as the QM region, the QM solvent at the edge of the solvent sphere is oriented to interact with additional solvent molecules rather than vacuum, creating a poor  electronic environment for these edge solvent molecules. In addition to point charges or a polarizable continuum improving the description of the edge solvent molecules, the structure of the solvent at the edge of the solvation sphere can also be relaxed to improve the imbalanced description obtained from only retaining the closest solvent molecules in the QM region.  \cite{Isborn2013, Lever2013} 

\begin{figure}
\centering
{\includegraphics[width=0.48\textwidth]{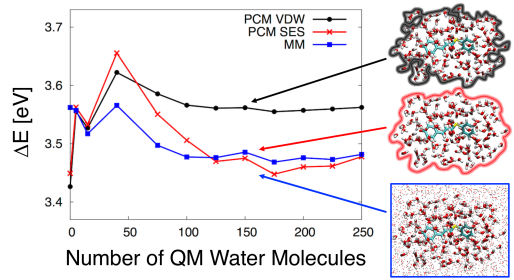}}
\caption{The computed excitation energy with PCM and MM solvent surrounding the QM solvent. The PCM employs either van der Waals (VDW) atomic radii for determining the molecular cavity or a solvent excluded surface (SES). Creating a cavity from VDW atomic radii leads to unphysical double counting of solvation effects from both the QM solvent and the continuum dielectric between QM solvent molecules. The solute is the anionic chromophore of photoactive yellow protein. More details can be found in Refs. \cite{Provorse2016} and \cite{Provorse_Long_2017}. Figure is adapted with permission from Ref. \cite{Provorse2016}. Copyright 2016 American Chemical Society.}
\label{fig:PCM_MM_ex_energy}
\end{figure}

For a large enough QM solvent region and a judicious choice of molecular cavity (more on this choice below), the polarization provided by fixed MM solvent point charges and the self-consistent polarization provided by a continuum model lead to the same excitation energy, see Figure \ref{fig:PCM_MM_ex_energy}.\cite{Provorse2016}  Obtaining the same value of the excitation energy using a 32 {\AA}ngstrom sphere of MM point charges and a polarizable model of the bulk solvent suggests that there is no need to include MM charges beyond this distance, for example via Ewald summation, in the computation of vertical excitation energies. The same may not be true for the computation of other properties, such as solvation energies and ionization potentials. The rate of the convergence of the excitation energy with size of QM region is also similar for both classical solvation models. Recent studies combining a polarizable MM  model with QM treatment of the environment provide improved convergence of excitation energies, allowing for a smaller QM region.\cite{Daday2015, Loco_2016, Dziedzic2016, Kratz2016, Nabo2017}

When combining QM solvent with a polarizable continuum model, careful consideration should be given to the choice of molecular cavity. The van der Waals atomic radii, which is the default radii in many electronic structure programs, may lead to spurious double counting of solvation interactions due to the surface charges appearing between QM solvent molecules rather than only outside of the QM region. The excess polarization provided with this choice of cavity will alter the excitation energies, as shown in Figure \ref{fig:PCM_MM_ex_energy}. The unphysical over-polarization can be removed by scaling the van der Waals radii to a larger value or by using a solvent-excluded surface. \cite{Provorse2016, Provorse_Long_2017} The use of a solvent-excluded surface is recommended for the calculation of vertical excitation energies when combining QM solvent with a continuum model because there is no need to choose the value for scaling the van der Waals radii. However, solving for the dielectric response with a solvent-excluded surface is often more computationally expensive than with a van der Waals surface, and a solvent-excluded surface can also lead to problematic geometry optimization convergence due to steep changes in the energy when changing the positions of the nuclei.

\begin{figure}
\centering
{\includegraphics[width=0.48\textwidth]{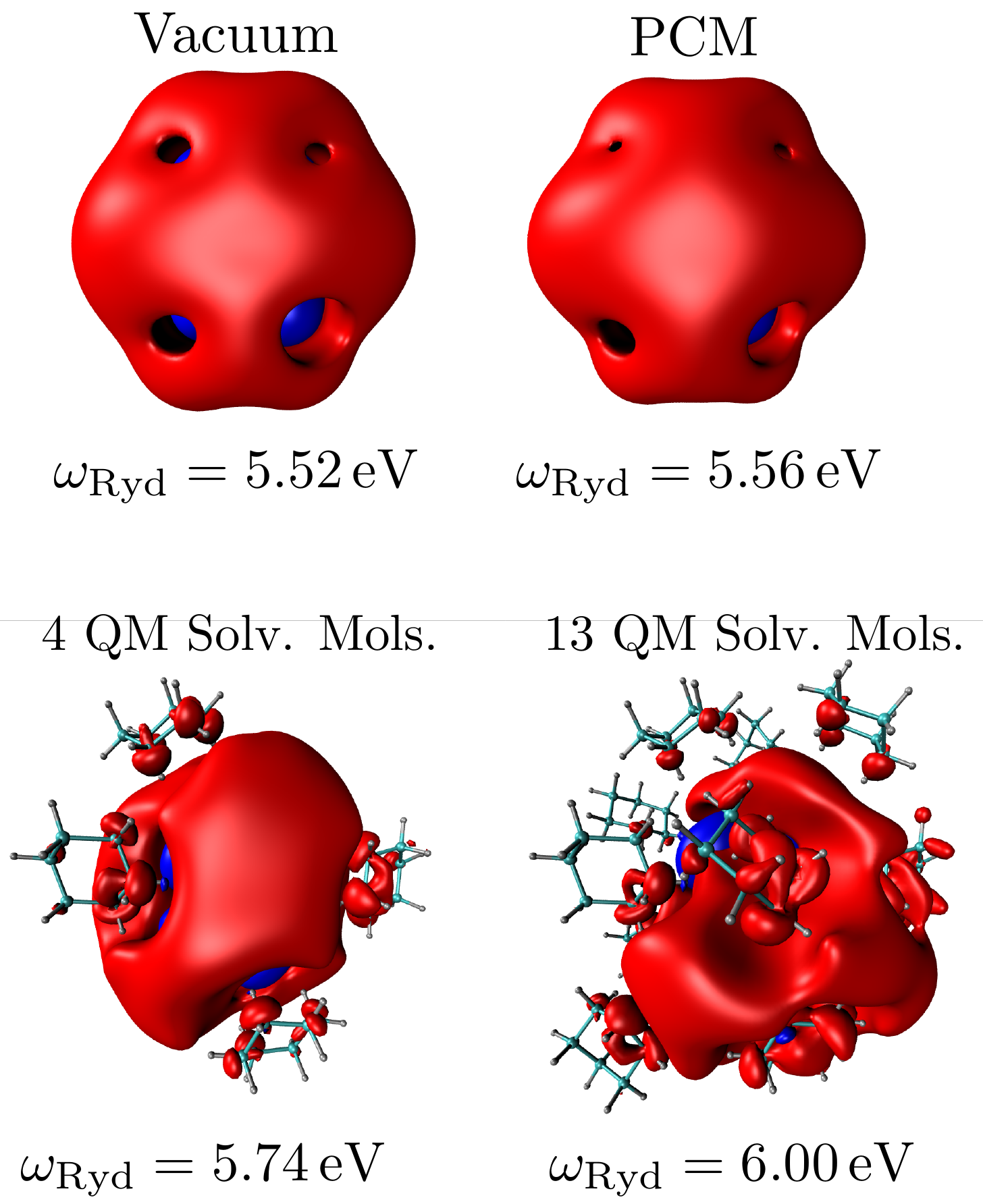}}
\caption{Electron-hole difference density plots for the 1 $^1A_u$ Rydberg state of naphthalene as computed in vacuum and in cyclohexane with different solvent representations. The electron density is plotted in red, the hole density in blue. An identical isodensity value is used in each plot. All calculations are carried out at the CAM-B3LYP/6-31++G** level of theory within the Tamm-Dancoff approximation using Terachem.}
\label{fig:rydberg_state_convergence}
\end{figure}

So far the discussion presented in this tutorial review has been focused on the computation of low-lying bright excitations that form the dominant contribution to the optical absorption spectra of solvated dyes. These states retain their relatively localized character in solution, with only small fractions of electron and hole density delocalizing onto the solvent through dynamic polarization effects\cite{Zuehlsdorff2016}. However, the choice of solvent representation can  have a strong influence on the character of the calculated excited state, as is most evident for Rydberg transitions. Here, we illustrate this point by computing the excitation energy of the 1 $^1A_u$ Rydberg state of naphthalene\cite{Peter2009}, both in vacuum and in cyclohexane, where the solvent environment is accounted for through a PCM or a mixed QM/MM solvent model. For the explicit solvent model, we focus on a single snapshot of naphthalene in a box of 427 cyclohexane molecules equilibrated using standard AMBER force fields, where the naphthalene chromophore is constrained to its ground state DFT-optimized structure. Two different explicit solvent representations are considered: One where the four nearest neighbor cyclohexane molecules are treated fully quantum mechanically and one where the QM region is extended to include the nearest 13 cyclohexane molecules. All molecules not within the QM region are included in the calculation as MM point charges. Plots of the electron-hole difference density, as well as the excitation energy of the Rydberg state in each solvent model can be found in Fig. \ref{fig:rydberg_state_convergence}. 

The character of the low-lying Rydberg state of naphthalene is almost identical in vacuum and in  a PCM, with only a small increase of excitation energy of 0.04~eV for the solvated system. However, as soon as an explicit solvent representation is introduced, the excitation energy of the Rydberg state increases significantly in energy, with the QM region containing 13 solvent molecules (corresponding to a total of 252 atoms in the QM region) leading to a blue shift of more than 0.4~eV when compared to the PCM results. The electron-hole difference density plot shows that the presence of the explicit solvent  strongly perturbs the delocalized electron state, an effect that cannot be reproduced with a PCM. Given that the representation of Rydberg states requires the use of significantly larger basis sets than needed for more localized valence states, systematic convergence tests of these states in solution with respect to the size of the QM region are computationally challenging. However, although the exact nature of these states in solution remains relatively unexplored, an explicit QM solvent representation is vital to capture the correct character and purely classical representations through a PCM or through MM point charges are insufficient to model Rydberg-type transitions in solution. 

\section{Excited State Methods}

After sampling solute-solvent configurations and choosing a solvent model, the remaining task is to compute the electronic excitation of the solute in the solvent environment. As in ground state calculations, the choice must be made between computational cost and accuracy. 

Wavefunction based excited state methods that include electron correlation such as equation of motion coupled cluster singles and doubles (EOM-CCSD),\cite{Stanton1993,Krylov2008} complete active space self-consistent field (CASSCF),  multi-reference configuration interation (MR-CI), as well as quantum Monte Carlo, have the potential to accurately simulate electronic excited states. Polarizable continuum models have been implemented with these excited state methods, see for example \cite{Floris2014, Ren2017}, but these techniques are often too computationally expensive to treat more than a few solvent molecules at the same level of theory as the solute. However, recent progress going beyond a polarizable continuum approach has been achieved. For example, quantum Monte Carlo has been interfaced with a polarizable MM method\cite{Guareschi2016} and used in conjunction with density functional theory (DFT) embedded potentials for modeling the excited states of small organic molecules in water.\cite{Daday2014} Also, projector based embedding was used to combine EOM-CCSD, considered the gold-standard method for simulating excited states, with a DFT-based potential for the environment, achieving excitation energies within 0.01 eV of the full EOM-CCSD results for acrolein solvated in water.\cite{Bennie2017} 

Linear response (LR)-based time-dependent DFT (TDDFT)\cite{Runge1984, Casida1995}, time-dependent Hartree-Fock (TDHF), and configuration interaction singles (CIS) methods are computationally affordable enough to include large QM regions of solvent around a molecule while computing vertical excitation energies. With the advent of linear-scaling TDDFT and electronic structure code that takes advantage of GPU parallel processing capabilities, large QM regions can be accommodated in the excited state calculation, up to a few thousand atoms. Solving the LR TDHF and TDDFT equations requires the functional derivative of the potential with respect to the electron density, therefore only terms in the potential that depend on the density will appear in the LR equations. The electrostatic embedding of MM point charges only affects the excitation energy through the polarization of the ground state electron density and these charges do not appear in the LR equations. However, there will be terms in the LR equations due to the solvent potential responding to the change in the electron density for polarizable continuum methods, for polarizable MM methods, and for the polarization component of fragment methods.  

As with ground state DFT, there are many options when choosing a density functional for computing excited states with TDDFT. Extensive benchmarking of density functionals has been performed for computing polarizable continuum based condensed-phase excitation energies \cite{Goerigk2009, Goerigk2010, Jacquemin_2012, Charaf-Eddin2013, Jacquemin2014, Guido_2015} and dipole moments.\cite{Guido2018} Although LR TDDFT is computationally affordable enough to include many solvent molecules in the QM region, the method must be used with caution with explicit QM solvent because the standard approximations employed with most density functionals will lead to charge-transfer transitions between solute and solvent that are spuriously low in energy, often lower in energy than the valence transitions of the solute.\cite{Bernasconi2003,Neugebauer2005, Lange2007, Isborn2013} Conventional local and semi-local functionals do not have the correct long-range spatial dependence to account for the Coulombic interaction between an excited electron and hole, and as a result the charge-transfer excitation energies are usually very close to the occupied-virtual orbital energy difference.\cite{Dreuw2003, Gritsenko2004, Maitra2017}

\begin{figure}
\centering
{\includegraphics[width=0.48\textwidth]{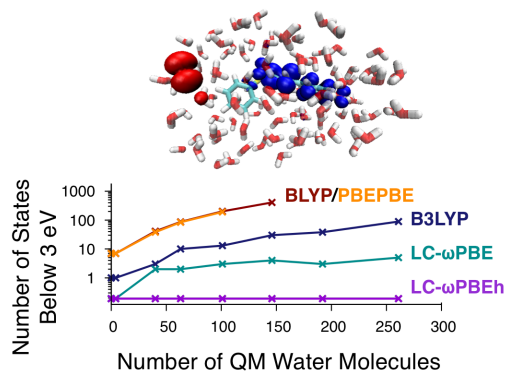}}
\caption{The TDDFT charge-transfer error manifests in solution with many low-energy transitions between the solvent and solute, as in the excited state density difference shown. These spuriously low-energy states occur with semi-local functionals (BLYP and PBEPBE), with hybrid functionals (B3LYP), and with long-range corrected hybrid functionals (LC-$\omega$PBE). Only with a sufficient amount of exact exchange at short range and full exact exchange at long range (LC-$\omega$PBEh) is the lowest energy transition a valence transition of the solute with all amounts of QM solvent. The solute is the anionic chromophore of photoactive yellow protein. Figure is adapted with permission from Ref. \cite{Isborn2013}. Copyright 2013 American Chemical Society}.
\label{fig:TDDFT_CT_Error}
\end{figure}

The short-range solute-solvent interactions can be modeled by extracting the closest solvent molecules around a solute from a larger-scale MD simulation. If the long-range electrostatic screening is neglected during this process, the semi-local DFT description of the orbitals of the solvent molecules often predicts an overly small band gap,\cite{Isborn2013, Lever2013} leading to many low-energy charge transfer transitions between the solvent and solute. The increase in the number of these low energy transitions with increasing amount of solvent is shown for the BLYP and PBEPBE functionals in Figure \ref{fig:TDDFT_CT_Error}. Beyond 150 solvent molecules, the band gap becomes too small using these functionals and the ground state calculation becomes unstable. The collapse of the band gap can be removed by correctly treating the long-range electrostatic effects of the bulk solvent on the QM solvent through MM charges or a PCM.\cite{Isborn2013, Lever2013}  However, spuriously low-energy charge transfer excitations between the solute and the solvent can remain problematic in local functionals, especially if the band gap of the solute is comparable to that of the solvent.  

Hybrid functionals include a fixed percentage of exact exchange in the potential, leading to wider band gaps, and therefore higher energy valence transitions and higher energy charge transfer transitions. Exact exchange provides the correct Coulombic stabilization of the electron-hole interaction, which for hybrid functionals will be scaled by the percentage of exact exchange, often in realm of 20-25\%. Because of this scaling, hybrid functionals, such as B3LYP, often still predict charge-transfer transitions to be too low in energy relative to valence transitions. Improved treatment of charge-transfer transitions is obtained with long-range corrected (LC) density functionals.\cite{Tawada2004, Rohrdanz2009} These functionals separate the Coulomb operator into long and short-range components, using the local functional to compute the exchange at short-range and exact orbital exchange for long-range,\cite{Leininger1997, Iikura2001, Vydrov2005} leading to the correct Coulombic potential between a transferring electron and hole. \cite{Tawada2004, Peach2006, Rohrdanz2009} For our test on a snapshot from an MD simulation,\cite{Isborn2013} the LC-$\omega$PBE functional with a range-separation parameter $\omega=0.2$ bohr$^{-1}$ predicted some low-energy charge-transfer transitions between solute and solvent, but adding some short-range exchange with the LC-$\omega$PBEh functional correctly predicted the valence transition of the chromophore to be the lowest energy transition for all amounts of explicit solvent, as shown in Figure \ref{fig:TDDFT_CT_Error}. 

One of the non-empirical ways to determine the range-separation parameter of LC functionals is through tuning of the ionization potential (IP) to match the negative energy of the highest occupied molecular orbital (HOMO).\cite{Stein2009, Baer2010,Kronik2012} Such IP tuning has been shown to give improved ionization potentials and excitation energies\cite{Jacquemin2014} over other functionals, although the size-dependence of the IP tuning procedure gives some cause for caution.\cite{K√∂rzd√∂rfer2014} A smaller range-separation parameter is predicted for increasing system size when using IP tuning, for both increasing molecular size and increasing amount of QM solvent, whereas improved agreement with $\lambda_{\textrm{max}}$ values is obtained by increasing the value of the range-separation parameter for molecules of increasing size.\cite{Garrett2014} IP tuning has been employed with polarizable continuum models, resulting in an overly small value for the range-separation parameter if using equilibrium solvation for both the neutral and cationic systems.\cite{deQueiroz2014, SosaVazquez2015, deQueiroz2015, Boruah2017} More reasonable values for the range-separation parameter are obtained for modeling ionization with a polarizable continuum model within the non-equilibrium TDDFT formulation.\cite{Rube≈°ov√°2017} 

\section{Computing lineshapes in solution}
In many instances, theoretical calculations of solvated dyes focus on the position of the absorption or emission band maximum. Properties such as solvatochromic shifts or Stokes shifts are then predicted by computing the vertical excitation energy for a single optimized configuration in implicit solvent and the resulting energy is directly compared to experimental $\lambda_{\textrm{max}}$ values of the band maximum. Although this approach is computationally appealing because it only requires the computation of excited states for a single representative structure in implicit solvent, thereby allowing for the use of highly accurate ab-initio quantum chemistry methods, it does not contain any information about the spectral lineshape. However, a significant number of excited state properties, such as the color of solvated dyes in solution, explicitly require an accurate prediction of the spectral lineshape. Even in situations when only the prediction of $\lambda_{\textrm{max}}$ values is of interest, such as in the prediction of solvatochromic shifts, estimates based on vertical excitations computed for a single representative structure can be unreliable, as they ignore a variety of effects such as temperature-, solvent- and lifetime broadening that can influence the position of experimental absorption maxima. Additionally, the spectra of many solvated dyes exhibit vibronic fine structure. For example, the double peak feature of Nile red in cyclohexane in Fig. \ref{fig:nile_red_experiment} is due to the vibronic fine structure of the bright S$_1$ transition \cite{Guido_2010}. Thus, a reliable prediction of spectral shapes requires the explicit consideration of the vibrational modes of the molecule on the ground and excited state potential energy surfaces.

In principle, it is highly desirable to develop accurate computational approaches that account for all effects that contribute to the spectral lineshape of a solvated chromophore. However, in practice, this task poses a substantial challenge. For the purpose of this perspective we will focus on the two contributions to the spectral shape most relevant for solvated dyes in solution, namely inhomogeneous solvent broadening and vibronic contributions.  
\begin{figure}
\centering
{\includegraphics[width=0.4\textwidth,angle=270]{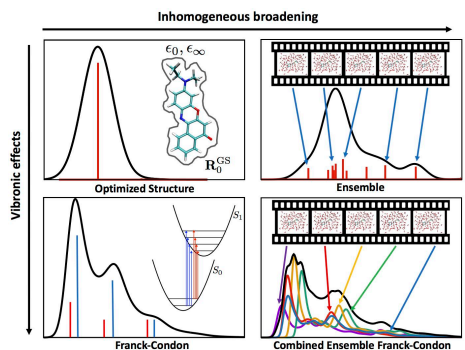}}
\caption{Schematic representation of four different approaches to computing the absorption lineshape of a solvated chromophore. To obtain accurate absorption lineshapes, it is desirable to explicitly account for both vibronic effects and inhomogeneous solvent broadening. }
\label{fig:different_lineshape_approaches}
\end{figure}

Methods to simulate absorption lineshapes are shown schematically in Figure \ref{fig:different_lineshape_approaches}. The most basic and computationally efficient approach computes a vertical excitation for the optimized ground state geometry of the solute in an implicit solvent model. In this approach, solvent effects are accounted for collectively through the polarization provided by the bulk solvent as determined by the static and dynamic dielectric constants. The absorption line shape consists of a phenomenological broadening function centered on the vertical excitation energy. 

A straightforward, albeit computationally considerably more expensive, way of going beyond this simple phenomenological lineshape is to account for inhomogeneous broadening through the sampling of solute-solvent conformations in the ensemble approach. Vertical excitation energies are computed for each configuration from the ensemble and are then convoluted with a broadening function before being summed into a spectrum, as in Fig.~\ref{fig:convergence_num_snapshots}. The ensemble approach has a number of appealing features, such as temperature effects arising classically from the MD sampling and the possibility of treating solvent polarization from first principles by including large parts of the solvent environment in the QM region of the excited state calculation. The approach has been successfully applied to predicting both solvatochromic shifts and colors of of a variety of solvated chromophores\cite{Zuehlsdorff2017,Malcioglu2011,deMitri2013}. However, spectra that are considerably too narrow compared to experimental results\cite{Isborn2012, Zuehlsdorff2018} or that are missing vibronic spectral features\cite{Tapavicza2016} can also result from the ensemble approach. Both of these failures can be ascribed to the neglect of vibronic effects.

To account for vibronic effects in the computation of absorption lineshapes it is necessary to go beyond a description of the nuclei as purely classical particles. 
Following the Franck-Condon principle, the overlap of the ground and excited state nuclear wavefunction defines the intensity of a vibronic transition.\cite{Hazra2003,Dierksen2005,Jankowiak2007,Santoro2008,Bloino2010,Santoro2016} Implementations of Franck-Condon vibronic spectra in electronic structure packages often rely on approximating the potential energy surfaces of the electronic states as harmonic around their minima\cite{Santoro2008,Santoro2016}, such that the vibrational wavefunctions and Franck-Condon overlaps are calculated from the Hessian computed at the two respective minima. Temperature effects are accounted for quantum mechanically by considering a Boltzmann distribution of initial vibrational states on the ground state potential energy surface\cite{Santoro2007,Santoro2007b}, producing a finite temperature Franck-Condon (FTFC) spectrum. Solvent polarization effects are included in FTFC spectra by computing the optimized ground and excited state geometries and vibrational modes within implicit solvent. To generate the spectral lineshape of a solvated chromophore from the Franck-Condon spectrum, the inhomogeneous solvent broadening is usually simulated by applying Gaussian broadening onto every transition in the FTFC spectrum computed in PCM, where the standard deviation of the Gaussian broadening function is either derived from the solvent reorganization energy in PCM\cite{Ferrer2011} or from MD sampling of solvent conformations around the chromophore frozen in its ground state structure.\cite{Zalesny2015,Cerezo2015}  The assumptions underlying this procedure are that the solvent broadening leads to the same Gaussian distribution around each vibronic transition and that the solvent motion is fully decoupled from the solute degrees of freedom, which are treated quantum mechanically in the Franck-Condon picture. 

The vibronic spectra computed with the FTFC approach can be rather sensitive to the choice of the DFT exchange-correlation functional\cite{Bednarska2017,Zuehlsdorff2018},  especially for systems where the excited state of interest has some intramolecular charge-transfer character. However, the vibrational reorganization energy can be used as a reliable metric for assessing the quality of a given exchange-correlation functional\cite{Bednarska2017b}, thus in principle allowing for the selection of a functional that most closely matches higher order quantum chemistry methods.

\begin{figure}
\centering
\includegraphics[width=0.48\textwidth]{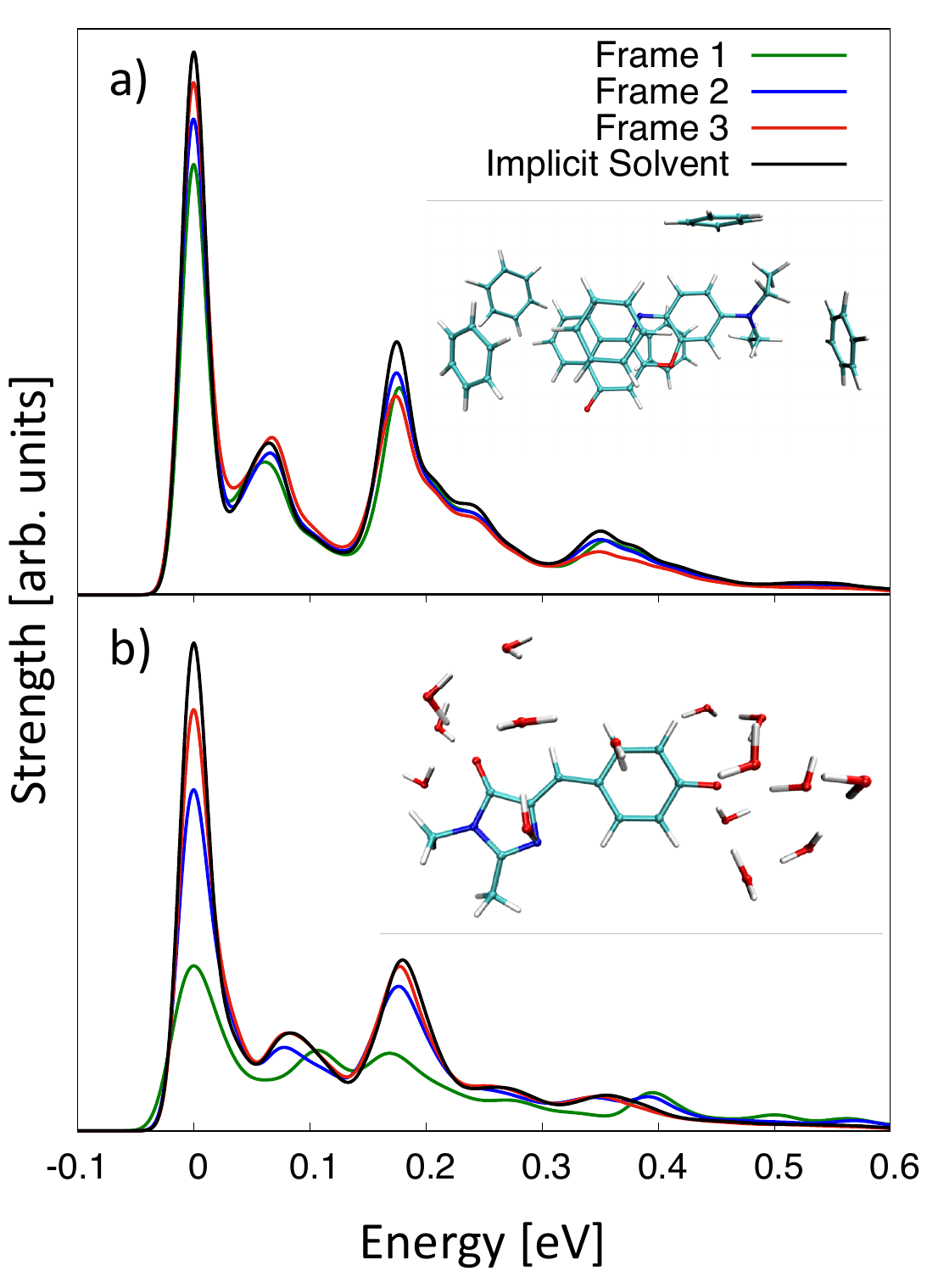}
\caption{Zero-temperature Franck-Condon spectra computed for a) Nile red in benzene and b) the GFP chromophore anion in water. Spectra are computed for three uncorrelated solute-solvent snapshots, where some nearest neighbor solvent molecules are explicitly included in the QM region and are kept frozen during the calculation of the vibronic transitions, whereas long-range electrostatic effects are accounted for using a PCM. The vibronic spectra computed in implicit solvent without any explicit solvent molecules in the QM region are displayed for comparison. This figure is adapted from Ref. \cite{Zuehlsdorff2018} with permission of AIP publishing. }
\label{fig:vibronic_comparison}
\end{figure}

The FTFC approach accounts for vibronic and temperature effects on the absorption line shape. However, the solvent effects are usually accounted for via a continuum approach, which can be problematic in systems with strong solute-solvent interactions. Figure \ref{fig:vibronic_comparison} shows zero-temperature FC vibronic spectra computed for Nile red in benzene and the GFP chromophore anion in water, where a few solvent molecules closest to the chromophores are explicitly included in the calculation. Three different snapshots of solute-solvent conformations are considered and the solvent molecules are kept fixed during the computation of optimized geometries and Hessians. Although these FC spectra do not account for any vibronic coupling between the solute and the solvent, the explicit solvent environment clearly influences the vibronic spectra of the chromphore if there are strong solute-solvent interactions. For Nile red in benzene the three vibronic spectra are in very close agreement with the spectrum obtained in implicit solvent. However, for the GFP chromophore anion in water, a system with much stronger solute-solvent interactions, there is strong variation between spectra computed in different frozen solvent configurations, suggesting that the vibronic spectra computed with a PCM neglects important explicit solvent effects. 

Another drawback of the FTFC approach is that the temperature-dependent inhomogeneous broadening does not arise from the interaction of a fully flexible solute with explicit solvent. Broadening derived from the PCM solvent reorganization energy neglects specific solute-solvent interactions such as hydrogen bonding. Broadening derived from the spread of vertical excitation energies for different solute-solvent conformations computed around the solute frozen in its ground state minimum neglects indirect solvent effects due to solvent-induced changes in the conformations of the chromophore. As a result of these neglected explicit solute-solvent interactions, recent studies\cite{Ferrer2011,Ferrer2014} have found that both inhomogeneous broadening approaches can yield spectra that are too narrow compared to experimental results, especially in systems with moderate to strong solute-solvent interactions. Thus, spectra computed within the ensemble approach may be too narrow due to the lack of vibronic effects, whereas spectra computed within the FTFC approach may be too narrow due to the lack of sampling solute-solvent configurations. 

To overcome the limitations of both the ensemble and the FTFC approaches, the two methods could potentially be combined into a single approach that accounts for both inhomogeneous broadening and vibronic effects without resorting to continuum models, shown schematically in Fig. \ref{fig:different_lineshape_approaches}. In this combined ensemble plus finite temperature Franck-Condon approach, the line shape of the absorption spectrum is generated by averaging over FTFC spectra calculated for $N$ uncorrelated snapshots of solute-solvent conformations, where the vibrational frequencies of the chromophore are computed inside a pocket of frozen solvent. In principle, such an approach allows for a rigorous computation of the influence of the explicit solvent environment of a given snapshot on the vibronic spectrum of the solute by including large parts of the solvent environment explicitly in the QM region used for the computation of the Franck-Condon spectra. Thus the resulting spectrum accounts account for inhomogeneous broadening and vibronic effects on equal footing. However, this rigorous approach comes at a very high computational cost as it requires the computation of the ground and excited state frequencies of the chromophore inside frozen QM solvent pockets for a large number of uncorrelated solute-solvent conformations. 

\begin{figure}
\centering
\includegraphics[width=0.48\textwidth]{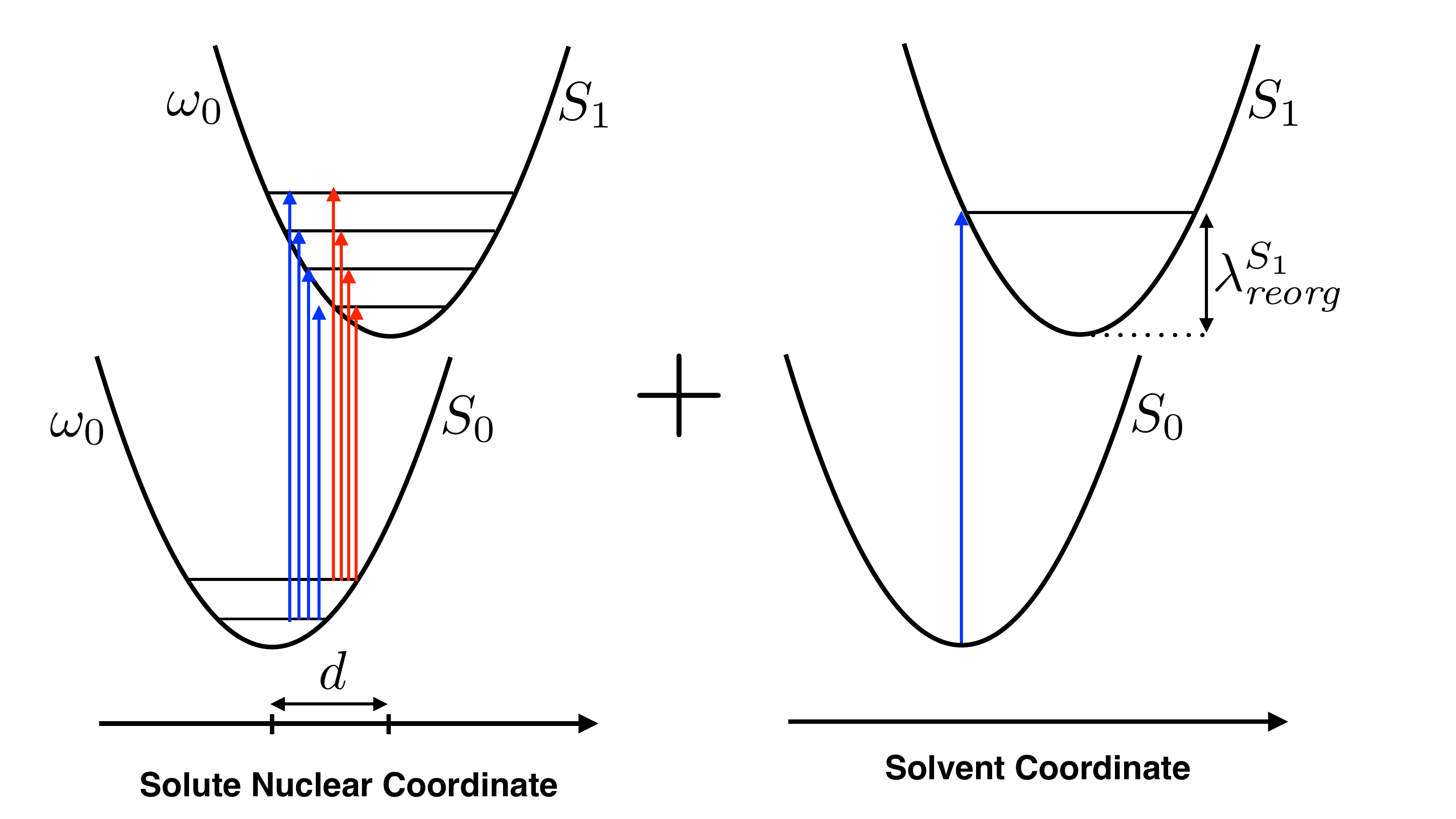} 
\caption{A simple model system for studying the absorption lineshape of solvated chromophores under different approximations. The model consists of an ensemble of identical displaced harmonic oscillators coupled to a classical solvent environment whose influence on the electronic excitation is described by the solvent reorganization energy $\lambda^{S_1}_{reorg}$.}
\label{fig:model_system}
\end{figure}

In order to avoid the high computational cost associated with computing Franck-Condon spectra for a large number of individual solute-solvent conformations, we have recently introduced an approximate way of combining the ensemble and the Franck-Condon approaches in two different temperature regimes, referred to as the ensemble plus zero-temperature Franck-Condon (E-ZTFC)\cite{Zuehlsdorff2018} approach. In the E-ZTFC approach, an effective average zero temperature Franck-Condon shape function is constructed by computing Franck-Condon spectra for a small number of solute-solvent configurations. The average Franck-Condon spectrum is then convoluted with the ensemble spectrum constructed from the vertical excitation energies computed for a large number of uncorrelated solute-solvent snapshots. The resulting spectrum accounts for all temperature effects of the chromophore and the solvent fully classically through the MD sampling, while introducing a zero temperature vibronic correction to the ensemble spectrum through the average Franck-Condon shape function. The approximate E-ZTFC approach significantly reduces the number of Franck-Condon spectra that need to be computed, such that constructing the final lineshape is not much more expensive than computing spectra in the ensemble approach. A drawback of the E-ZTFC approach is that it includes partial double counting of the ground state vibrational motion of the chromophore, which is accounted for both in the zero temperature Franck-Condon spectrum and the conformations sampled by the MD. 

The effect of the double counting on the spectral shape can be assessed by studying a simple model system of an ensemble of identical displaced harmonic oscillators coupled to a classical solvent environment (see Fig. \ref{fig:model_system} for a schematic of the model system). Because this simple model is fully harmonic, the FTFC approach is exact in this system, whereas the ensemble and the E-ZTFC approaches are approximate for the model system. Thus, comparing the E-ZTFC lineshape with the FTFC lineshape allows us to quantify the influence of the double-counting of vibrational degrees of freedom on the computed spectrum. However, it should be stressed that in realistic systems with complex solute-solvent interactions and anharmonic degrees of freedom, the FTFC approach will not be exact and the anharmonic regions of the potential energy surface can be sampled within the ensemble. In these realistic systems, the combined E-ZTFC approach has the advantage of fully accounting for solute-solvent interactions when computing inhomogeneous broadening effects. 

\begin{figure}
\centering
\includegraphics[width=0.48\textwidth]{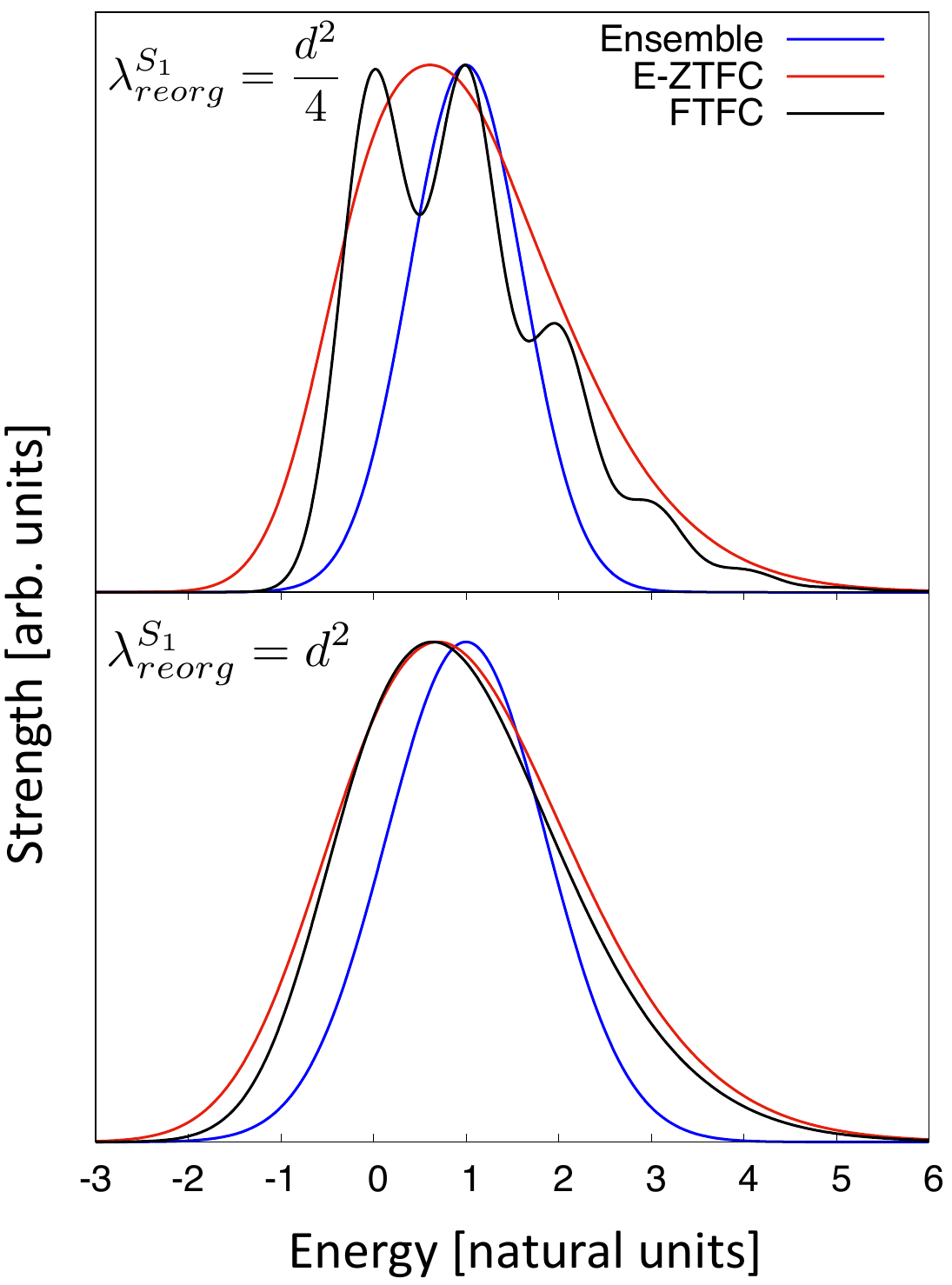} 
\caption{A comparison between the absorption spectra generated in the ensemble, the FTFC, and the hybrid E-ZTFC approaches for the simple model system introduced in Fig. \ref{fig:model_system}. The energy scale is set at $\frac{\hbar\omega_0}{k_B T}=0.125$, approximately corresponding to a C-C double bond oscillation at room temperature. Results are presented for two different solute-solvent interaction strengths, with $\lambda_{reorg}^{S_1}=\frac{d^2}{4}$ and $\lambda_{reorg}^{S_1}=d^2$ (expressed in natural units). This figure is reprinted from Ref. \cite{Zuehlsdorff2018} with permission from AIP publishing.}
\label{fig:model_system_results}
\end{figure}

Fig. \ref{fig:model_system_results} shows the resulting spectra of the classical ensemble, the fully quantum mechanical FTFC, and the mixed E-ZTFC approaches for the model system with parameters chosen to approximate a C-C double bond oscillation at room temperature and two different solute-solvent interaction strengths modeled through the solvent reorganization energy $\lambda_{reorg}^{S_1}$. For comparatively weak solute solvent interactions, the FTFC approach retains a considerable degree of vibronic fine structure that washes out in the combined E-ZTFC approach due to the double counting of solute vibrational degrees of freedom. However, for stronger solute-solvent interactions, the E-ZTFC approach produces results that are almost identical to the exact FTFC spectrum, while providing a significant improvement over the line shape as computed in the ensemble approach.

\begin{figure}
\centering
{\includegraphics[width=0.48\textwidth]{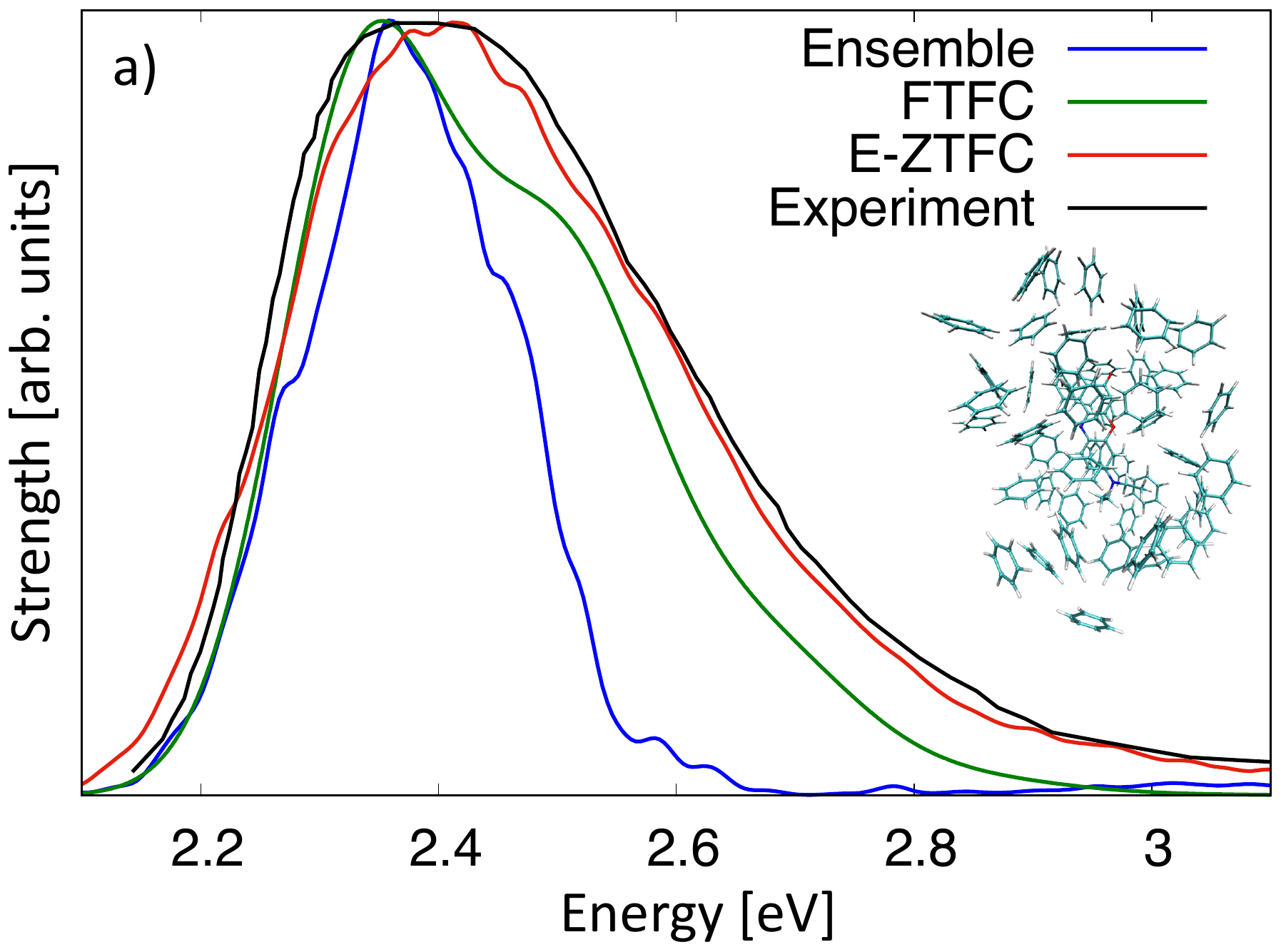}} \\
{\includegraphics[width=0.48\textwidth]{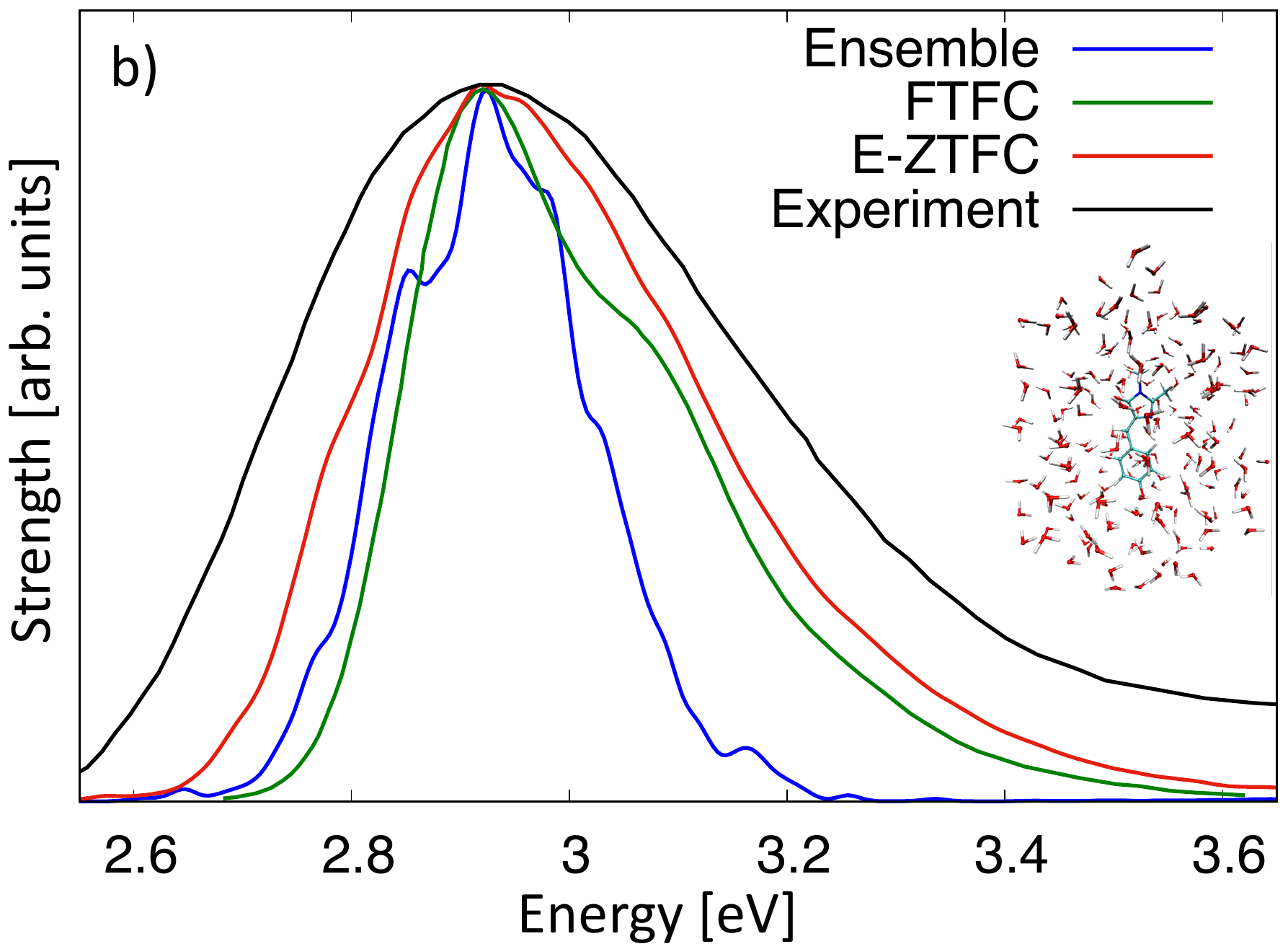}}
\caption{Spectra computed in the ensemble and the E-ZTFC approaches for a) Nile red in benzene and b) the GFP chromophore anion in water. For comparison, we also show the spectrum obtained from a pure FTFC approach, with solvent broadening derived from an explicit solvent representation around a solute frozen at its ground state minimum (benzene) or directly from the solvent reorganization energy calculated in a PCM (GFP chromophore anion); the GFP chromophore FTFC results are from Avila Ferrer \emph{et al}\cite{Ferrer2014}). Experimental spectra are from Ref. \cite{Davis_1966} (benzene) and Ref.\cite{Nielsen2001} (GFP chromophore anion) respectively. This figure is adapted from Ref. \cite{Zuehlsdorff2018} with permission of AIP publishing.}
\label{fig:lineshape_spectra_comparison}
\end{figure}

To assess the quality of the lineshape computed in the E-ZTFC approach in realistic systems, we have recently simulated the absorption spectra of a number of small solvated dyes\cite{Zuehlsdorff2018}. Fig. \ref{fig:lineshape_spectra_comparison} shows the computed lineshapes for the ensemble, FTFC, and E-ZTFC  approaches for Nile red in benzene and the GFP chromophore anion in water along with experimental spectra. The E-ZTFC lineshape is significantly improved compared to the ensemble lineshape, as well as the FTFC lineshape with solvent broadening derived from a PCM model. The ensemble approach yields spectra that are much too narrow and that lack the vibronic tail at high energy, in agreement with the results for the simple model system of displaced harmonic oscillators (see Fig. \ref{fig:model_system_results}). The FTFC spectra retain some vibronic fine structure, contrary to the smooth experimental results, and are too narrow, suggesting that the amount of inhomogeneous broadening provided by the environment is underestimated. 

The combined E-ZTFC approach appears to be more successful in accounting for inhomogeneous broadening in these systems, but the spectrum generated for the GFP chromophore anion in water is still significantly too narrow compared to the experimental spectrum. Potential sources for the missing width could be the neglect of counter-ions in the MD sampling of solute-solvent configurations, the lack of nuclear quantum effects in the sampling, or the missing homogeneous broadening due to the finite lifetime of the excited state. Furthermore, as shown in Fig. \ref{fig:vibronic_comparison}, the vibronic fine structure of the GFP chromophore anion shows strong variations with changes in the solvent environment, suggesting that approximating the vibronic fine structure through a single average shape function (or a vibronic spectrum computed in a PCM) may be a poor approximation. 

The E-ZTFC approach provides a promising new way to account for both vibronic and inhomogeneous broadening effects in the spectra of solvated chromophores, with a computational cost that is not significantly higher than that of the pure ensemble approach. The ensemble sampling produces temperature-dependent inhomogeneous broadening due to explicit solute-solvent interactions, as well as a sampling anharmonic regions of the potential energy surface. These anharmonic contributions to the spectra are not accounted for by Franck-Condon spectra constructed in the harmonic approximation. Although the more rigorous combined ensemble plus finite-temperature Franck-Condon approach described schematically in Fig. \ref{fig:different_lineshape_approaches} is free of any double counting of vibrational degrees of freedom and computes a unique Franck-Condon spectrum for each frozen solvent environment, its high computational cost places limitations on its practical use for computing spectra of solvated dyes. The computationally much cheaper E-ZTFC method has the potential to capture the most important vibronic and inhomogeneous broadening effects in solvated dyes, allowing for the computation of accurate absorption spectra in a large variety of systems. 

\section{Successes and Challenges}
The development of novel algorithms for large-scale electronic structure calculations, as well the improved availability of computational resources, have led to a steady increase in recent years of computational studies of excited states of molecules in solution that go beyond the computation of a single vertical excitation energy in implicit solvent. These studies include the computation of solvatochromic shifts, Stokes shifts, as well as full absorption and emission lineshapes taking into account vibronic effects. The computation of vertical excitation energies with explicit solvent models is now relatively well established. Recent work in the field has defined robust computational protocols regarding the amount of QM solvent necessary to capture polarization effects from first principles, as well as how to best treat long-range electrostatic contributions. As algorithms and new computational capabilities are developed to speed up the calculation of excited states using highly accurate wavefunction methods, these methods can employ similar protocols that have been pioneered with the TDDFT methodology. 

One of the main challenges in the computation of an accurate ensemble of vertical excitation energies lies in the efficient MD sampling of a large number of uncorrelated solute-solvent conformations. If the sampling is carried out using MM force fields, it is often necessary to carefully re-parametrize certain force field parameters of the chromophore\cite{Zuehlsdorff2017,Zuehlsdorff2018}, as the choice of force field can have a strong influence on computed spectra\cite{Cerezo2016}. To improve the accuracy of a spectrum generated from MM configurations, the computed excitation energies can also be reweighted using the difference between the MM and QM energies, as is often done for the computation of enzymatic reaction barriers.\cite{Cooper2014} Sampling with the chromophore treated fully quantum mechanically can alleviate this issue, but the choice of solvent force field can lead to significant differences in the sampled solvent conformations in the first solvation shell\cite{Cerezo2015}. In systems with strong solute-solvent interactions, such as for anionic chromophores in water, contributions to the spectrum from highly shared protons between the solute and the solvent are generally missed by MM force field treatments. In principle, these effects can be rigorously captured by obtaining the conformations directly from ab-initio MD or ab-initio path integral MD, but these computationally expensive approaches place limits on the simulation cell sizes, as well as the number of uncorrelated snapshots that can be obtained. An extreme challenge would be reproducing the experimental spectrum for molecules in different protonation states from an ensemble of vertical excitation energies. Accurate sampling for these systems would likely require ab initio path integral MD with a simulation box large enough to include counter ions at the correct concentration and with long enough timescales to see proton transfer and equilibration. Because of computational limitations, this is not achievable in practice, and the influence of counter ions and box size on lineshape remains relatively unexplored.

Explicit solvent models, especially those based on including large parts of the solvent environment explicitly in the computation of vertical excitation energies, have been successful in predicting spectral properties in situations where implicit solvent approaches are known to fail, such as in systems with strong solute-solvent interactions. This is particularly apparent when attempting to predict relative solvatochromic shifts of a chromophore in solvents with similar dielectric properties but strong differences in solute-solvent interaction strengths. For example, implicit solvent approaches are unable to reproduce the significant difference in solvatochromic shift between Nile red in acetone and Nile red in ethanol, whereas an explicit solvent approach yields results in close agreement with experiment\cite{Zuehlsdorff2017}. Another study showed that implicit solvent approaches are unable to reproduce the strong solvatochromic shift of N,N-diethyl-4-nitroaniline in water\cite{Eilmes2014}, whereas explicit solvent yields much improved results. Explicit solvent models can also lead to improved solvatochromic shifts in non-polar solvents, where the small dielectric constant of the solvent produces small shifts relative to vacuum calculations for implicit solvent approaches. \cite{Zuehlsdorff2017,Zuehlsdorff_2015}

Beyond the computation of solvatochromic shifts, the ensemble approach based on an explicit treatment of the solvent environment has also been shown to yield good spectral lineshapes. For flexible chromophores with strong solute-solvent interactions, spectra computed from an ensemble of vertical excitation energies computed for independent solute-solvent snapshots are often in good agreement with experiment. For example, previous studies of cyanin in water\cite{Malcioglu2011,Ge2015} found that absorption spectra computed with an explicit solvent model reproduce experimentally observed colors. Recent studies demonstrated good agreement between experimental absorption and emission spectra and calculated ensemble spectra for oxyluciferin and its analogues\cite{Hiyama2017,Irepa2018}. In these systems, the spectral lineshape is dominated by solute-solvent interactions and low frequency motions of the chromophore, such as twisting around a dihedral angle, which often correspond to anharmonic regions of the potential energy surface. In systems where contributions from configurations on  anharmonic areas of the potential energy surface have a much stronger spectral contribution than vibronic effects, an ensemble approach based on explicit solvent representation and vertical excitation energies can yield accurate lineshapes. 

For relatively rigid chromophores in weakly interacting solvent, vibronic fine structure may dominate the spectral lineshape, leading the ensemble approach to fail. In these systems, the FTFC approach to computing lineshapes may be successful in predicting absorption and emission spectra. A general overview of the field can be found in a recent review\cite{Santoro2016}. Apart from yielding robust lineshapes for rigid systems, the FTFC approach can also account for Herzberg-Teller effects, allowing for the prediction of spectral lineshapes in transitions that have a very weak oscillator strength in vertical excitation models\cite{Ferrer2013,Baiardi2013,Santoro2011}. 

The main difficulty faced by the FTFC approach is the common reliance on the harmonic approximation for computing the vibrational wavefunctions of the ground- and excited state potential energy surface. This is likely a good approximation for rigid chromophores where vibronic contributions are due to high frequency modes, but the harmonic approximation starts breaking down for semi-flexible chromophores. If only a few low frequency modes, such as torsional rotations of rigid parts of the chromophore, are expected to contribute to the spectrum, these modes can be treated classically while retaining a full FTFC treatment of other modes\cite{Ferrer2014,Zalesny2015,Improta2014}. However, in situations were several moderately anharmonic modes contribute to a spectrum, a clear division into anharmonic classical modes and quantum harmonic modes treated with FTFC may no longer be accurate. Another challenge for the FTFC approach is that the vibronic transitions are usually computed for a structure optimized in continuum solvent, which will likely predict poor spectra in the same systems where continuum solvent approaches fail for vertical excitation energies, namely when attempting to compute relative solvatochromic shifts between solvents with similar dielectric properties but different solute-solvent interaction strengths. 

The E-ZTFC approach outlined in Ref. \cite{Zuehlsdorff2018} bridges the gap between the  ensemble approach that is accurate for flexible chromophores and the FTFC approach that performs well for rigid systems. The E-ZTFC method provides improved absorption lineshapes for Nile red and the GFP chromophore compared to the ensemble and FTFC approaches. The  improvements originate from the ensemble sampling of solute-solvent conformations while also including the direct influence of explicit solvent on the vibronic contributions to the spectra.
However, a variety of open questions and challenges remain for this approach. For example, although the amount of explicit QM solvent necessary to converge vertical excitation energies has been the focus of a number of studies, systematic tests regarding the influence of the amount of explicit QM solvent on computed Franck-Condon spectra are lacking. Furthermore, in systems with strong solute-solvent interactions, the Franck-Condon spectra computed in different frozen solvent environments are found to deviate significantly from each other, calling into question the use of a single vibronic shape function to account for the vibronic fine structure. In principle, these questions can be addressed by comparing the E-ZTFC approach to the more rigorous ensemble plus finite-temperature Franck-Condon approach based on a spectral average of a large number of independent Franck-Condon spectra. 

Calculating vibronic spectra within a Franck-Condon approach decouples the vibrational degrees of freedom of the solvent and the electronic degrees of freedom of the solute. This formalism assumes that only the solute's vibrational modes couple to the vertical excitations to produce the vibronic fine structure of the system, with the solvent providing electronic polarization and inhomogeneous broadening.  This approximation likely begins to break down for systems with very strong solute-solvent interactions, such as when proton sharing occurs between the solute and a protic solvent. Ideally, all motion of the solute and the solvent should be treated on the same footing, allowing for the coupling of all vibrational modes in the system to the electronic excitations. In the static frameworks based on uncorrelated MD snapshots presented in this work, such a full coupling is difficult to achieve, and therefore it becomes necessary in the future to consider fully dynamic approaches to lineshape theory. These dynamic formalisms model the interaction of an electronic system with a fluctuating environment through a spectral density of system-bath coupling that can be extracted from the energy gap autocorrelation function of the system evolving in equilibrium\cite{Mukamel1995}. The approach is often used to describe energy transfer in multi-chromophore systems,\cite{Chandrasekaran2015} but can also be applied to compute the lineshape of single chromophores in a complex environment\cite{Loco2018}. The advantage of a dynamic approach is that the motion of both the chromophore and the environment is coupled to the electronic degrees of freedom at the desired temperature. The approach is in general much more computationally demanding (many thousands of vertical excitation energy calculations are required) than the static approaches described in this work, but can potentially serve as an effective way to benchmark more approximate methods, highlighting situations where the static approximations break down.

\section{Acknowledgments}
CI and TZ have been supported by the Department of Energy, Offce of Basic Energy Sciences CTC and CPIMS programs, under Award Number DE-SC0014437.

\section{Biographies}
Tim Zuehlsdorff obtained his BSc in Theoretical Physics and Ph.D. in Theory and Simulation of Materials at Imperial College London. He worked as a postdoctoral researcher in the Theory of Condensed Matter Group under Professor Mike Payne at the University of Cambridge before joining Christine Isborn's group as a postdoctoral researcher at the University of California Merced in 2016. His research is focused on the modeling of electronic excitations in complex environments, as well as the development of linear-scaling electronic structure methods.  
\\
\\
Christine Isborn earned her B.S. in Chemistry and Physics at the University of San Francisco and her Ph.D. in Chemistry at the University of Washington, Seattle. She worked as postdoctoral researcher with Todd Martinez at Stanford University before becoming a professor at the University of California in Merced in 2012. Her research interests involve understanding how molecules interact with light and how molecular environments, such as solvent, ions, or interfaces, tune this light-matter interaction.

%

\end{document}